\documentclass[prl,twocolumn,showpacs,preprintnumbers,amsmath,amssymb]{revtex4}

\usepackage{natbib}
\usepackage{bibentry}
\usepackage{graphicx}
\usepackage{dcolumn}
\usepackage{bm}
\newcommand{\bra}[1]{\langle#1|}

\newcommand{\ket}[1]{|#1\rangle}

\begin{document}

\preprint{APS/123-QED}

\title{Non-Markovian effects on the dynamics of entanglement}

\author{B. Bellomo, R. Lo Franco, and G. Compagno}

\affiliation{Dipartimento di Scienze Fisiche ed Astronomiche,
Universit\`{a} di Palermo, via Archirafi 36, 90123 Palermo, Italy}

\date{\today}

\begin{abstract}
A procedure that allows to obtain the dynamics of $N$ independent
bodies each locally interacting with its own reservoir is
presented. It relies on the knowledge of single body dynamics and
it is valid for any form of environment noise. It is then applied
to the study of non-Markovian dynamics of two independent qubits,
each locally interacting with a zero temperature reservoir. It is
shown that, although no interaction is present or mediated between
the qubits, there is a revival of their entanglement, after a
finite period of time of its complete disappearance.
\end{abstract}


\maketitle

Entanglement is relevant to different fundamental
\cite{bell,bennett} aspects of quantum theory and practical
aspects of quantum information processing \cite{niels}. Recently
much interest has arisen in the evolution of the joint
entanglement of a pair of qubits exposed to local noisy
environments. The reason is related to the discovery by Yu and
Eberly \cite{yu1} that for this system, rather surprisingly, the
Markovian dynamics of the joint qubits entanglement and single
qubit decoherence may be rather different. The aspect that has
mostly drawn attention is the possibility of a complete
disappearance of entanglement at finite times. The occurrence of
this phenomenon, termed "entanglement sudden death" (ESD), has
been shown in a quantum optics experiment \cite{almeida}. The
intrinsic interest and potential importance of ESD, for example in
the application range of quantum error correction methods, has
lead to a flow of analysis that study its appearance under
different circumstances
\cite{yu2,santos,diosi,carv1,tanas,sun,hamdou,cirone}.

Disentanglement is related to the birth of body-environment
correlations. It is therefore of interest to investigate the role
played on its evolution by non-Markovian effects. In fact,
although Markovian dynamics includes a level of back-reaction, it
neglects the entanglement that arises between bodies and bath
modes during the evolution. Although some work has treated of this
aspect \cite{ban1,ban2,kuanliu,yon,glendinning}, it should be
considered an attractive theoretical challenge to extend the
results obtained under various conditions in the Markovian regime
to the non-Markovian case \cite{yu5}. The aim of this letter is to
address this point first by adopting a procedure to obtain the
dynamics of $N$ independent bodies, locally interacting with
reservoirs and without restriction on the nature of environmental
noise, if the single body dynamics is known. We shall then use
this approach to explicitly investigate the entanglement dynamics
of two qubits locally interacting with a zero temperature
non-Markovian environment.

To describe the method we consider a system formed by two
non-interacting parts $\tilde{A},\tilde{B}$, each part consisting
of a qubit $S=A,B$ locally interacting respectively with a
reservoir $R_S=R_A,R_B$. Each qubit and the corresponding
reservoir are initially considered independent. For each part, the
reduced density matrix evolution for the single qubit $S=A,B$ is
given by
\begin{equation}\label{reducedrho}
\hat{\rho}^{S}(t)=\textrm{Tr}_{R_S}\left\{\hat{U}^{\tilde{S}}(t)\hat{\rho}^{S}(0)
\otimes\hat{\rho}^{R_S}(0)\hat{U}^{\tilde{S}\dag}(t)\right\},
\end{equation}
where the trace is over the reservoir $R_S$ degrees of freedom and
$\hat{U}^{\tilde{S}}(t)$ is the time evolution operator for the
part $\tilde{S}$. In terms of the Kraus operators
$W_{\alpha,\beta}^{S}(t)$, the former equation becomes
\cite{kraus}
\begin{equation}\label{krausrep}
\hat{\rho}^{S}(t)=\sum_{\alpha\beta}W_{\alpha\beta}^{S}(t)\hat{\rho}^{S}(0)W_{\alpha\beta}^{\dag
\,S}(t).
\end{equation}
The assumption of independent parts implies that the time
evolution operator $\hat{U}^{\tilde{T}}(t)$ of the complete system
$\tilde{T}=\tilde{A}+\tilde{B}$ factorizes as
$\hat{U}^{\tilde{T}}(t)=\hat{U}^{\tilde{A}}(t)\otimes
\hat{U}^{\tilde{B}}(t)$. It follows that the Kraus representation
of the reduced density matrix for the two-qubit system $T=A+B$
reads like
\begin{equation}\label{kraustotal}
\hat{\rho}^{T}(t)=\sum_{\alpha\beta}\sum_{\gamma
\delta}W_{\alpha\beta}^{A}(t)W_{\gamma\delta}^{B}(t)
\hat{\rho}^{T}(0)W_{\alpha\beta}^{\dag
\,A}(t)W_{\gamma\delta}^{\dag \,B}(t).
\end{equation}
Given the basis $\{\ket{0},\ket{1}\}$ for each qubit, inserting
unity operators $I=\ket{0}\bra{0}+\ket{1}\bra{1}$ between Kraus
operators and density matrices in Eq.~({\ref{krausrep}}), it
follows that the dynamics of each qubit has the form
\begin{equation}\label{singleevo}
\rho^{A}_{ii'}(t)=\sum_{ll'}A_{ii'}^{ll'}(t)\rho^{A}_{ll'}(0),
\,\rho^{B}_{jj'}(t)=\sum_{mm'}B_{jj'}^{mm'}(t) \rho^{B}_{mm'}(0).
\end{equation}
Adopting the same procedure for $\hat{\rho}^{T}$ in the form of
Eq.~({\ref{kraustotal}}), the dynamics of the two-qubit system is
given by
\begin{equation}\label{totalevo}
\rho^{T}_{ii',jj'}(t)=\sum_{ll',mm'}A_{ii'}^{ll'}(t)
B_{jj'}^{mm'}(t)  \rho^{T}_{ll',mm'}(0),
\end{equation}
where the indexes $i,j,l,m=0,1$. Eqs.~(\ref{singleevo}),
(\ref{totalevo}) clearly show that the dynamics of two-qubit
density matrix elements can be obtained by knowing that of the
single qubit. The validity of above procedure can be
straightforwardly extended to any multipartite and multilevel
system (qudit), provided that the different parts, ``qudit +
reservoir'', are independent.

We now apply the results obtained above to study non-Markovian
effects on the entanglement dynamics of two qubits, each
interacting only and independently with its local environment. To
this aim we shall consider the single ``qubit+reservoir''
Hamiltonian given by
\begin{equation}\label{Hamiltonian}
H=\omega_0 \sigma_+\sigma_-+\sum_k \omega_k b_k^\dag b_k+(\sigma_+B
+\sigma_- B^\dag),
\end{equation}
with $B=\sum_k g_k b_k$, where $\omega_0$ is the transition
frequency of the two-level system (qubit) and $\sigma_ \pm$ are
the system raising and lowering operators while the index $k$
labels the field modes of the reservoir with frequencies
$\omega_k$, $b_k^\dag $, $b_k $ are the modes creation and
annihilation operators and $g_k$ the coupling constants. The
Hamiltonian of Eq.~(\ref{Hamiltonian}) may describe various
systems as for example a qubit formed by an exciton in a potential
well environment. However to fix our ideas we shall take it to
represent a qubit formed by the excited and ground electronic
state of a two-level atom interacting with the reservoir formed by
the quantized modes of a high-$Q$ cavity. At zero temperature,
this Hamiltonian represents one of the few open quantum systems
amenable to an exact solution \cite{garraway1997}. The dynamics of
qubit $S$ is known to be described by the reduced density matrix
\cite{petru,maniscalco}
\begin{equation}\label{roA}
\hat{\rho}^S(t)=\left(%
\begin{array}{cc}
\rho^S_{11}(0)P_t  & \rho^S_{10}(0)\sqrt{P_t}\\\\
\rho^S_{01}(0)\sqrt{P_t}  & \rho^S_{00}(0)+ \rho^S_{11}(0)(1-P_t) \\
\end{array}\right),
\end{equation}
where the function $P_t$ obeys the differential equation
\begin{equation}\label{equforp}
\dot{P}_t =-\int_0^t \mathrm{d} t_1 f(t-t_1)P_{t_1}\,,
\end{equation}
and the correlation function $f(t-t_1)$ is related to the spectral
density $J(\omega)$ of the reservoir by
\begin{equation}\label{corrfunc}
f(t-t_1)=\int \mathrm{d} \omega J(\omega)\exp
[i(\omega_0-\omega)(t-t_1)]\,.
\end{equation}
The exact form of $P_t$ thus depends on the particular choice for
the spectral density of the reservoir. Because the Hamiltonian of
Eq.~(\ref{Hamiltonian}) represents a model for the damping of an
atom in a cavity, we consider then the case of a single excitation
in the atom-cavity system. For the effective spectral density
$J(\omega)$, we take the spectral distribution of an
electromagnetic field inside an imperfect cavity supporting the
mode $\omega_0$, resulting from the combination of the reservoir
spectrum and the system reservoir coupling with $\gamma_0$ related
to the microscopic system-reservoir coupling constant, of the form
\cite{petru}
\begin{equation}\label{spectraldensity}
J(\omega)=\frac{1}{2 \pi}\frac{\gamma_0
\lambda^2}{(\omega_0-\omega)^2+\lambda^2}\,.
\end{equation}
The correlation function (\ref{corrfunc}) corresponding to the
spectral density of Eq.~(\ref{spectraldensity}) has an exponential
form with $\lambda$ as decay rate. The parameter $\lambda$,
defining the spectral width of the coupling, is then connected to
the reservoir correlation time $\tau_B$ by the relation $\tau_B
\approx \lambda^{-1}$. On the other hand the parameter $\gamma_0$
can be shown to be related to the decay of the excited state of
the atom in the Markovian limit of flat spectrum. The relaxation
time scale $\tau_R$ over which the state of the system changes is
then related to $\gamma_0$ by $\tau_R \approx \gamma_0^{-1}$.

Using the spectral density of Eq.~(\ref{spectraldensity}) in the
correlation function of Eq.~(\ref{corrfunc}), in the subsequent
analysis of the function $P_t$ of Eq.~(\ref{equforp}), typically a
weak and a strong coupling regime can be distinguished. For weak
regime we mean the case $\gamma_0<\lambda/2$, that is $\tau_R> 2
\tau_B$. In this regime the relaxation time is greater than the
reservoir correlation time and the behaviour of $P_t$ is
essentially a Markovian exponential decay controlled by
$\gamma_0$. In the strong coupling regime, that is for
$\gamma_0>\lambda/2$, or $\tau_R< 2 \tau_B$, the reservoir
correlation time is greater or of the same order of the relaxation
time and non-Markovian effects become relevant. For this reason we
are interested in this regime and we shall mainly limit our
considerations to this case. Within this regime, the function
$P_t$ assumes  the form \cite{petru,maniscalco}
\begin{equation}\label{p1}
P_t=\mathrm{e}^{-\lambda t}\left[ \cos \left(\frac{d
t}{2}\right)+\frac{\lambda}{d}\sin \left(\frac{d t}{2}\right)
\right]^2,
\end{equation}
where $d=\sqrt{2\gamma_0 \lambda-\lambda^2}$. $P_t$ presents
oscillations describing the fact that the decay of the atom
excited state is induced by the coherent processes between the
system and the reservoir. In particular he function $P_t$ has
discrete zeros at $t_n=2\left[n\pi-\arctan (d/\lambda )\right]/d$,
with $\textrm{$n$ integer}$. We note that the solution in the weak
coupling regime can be obtained by the former one simply
substituting the harmonic functions with the corresponding
hyperbolic ones and $d$ with $i d$.

Now we are ready to use, following the procedure described before,
the evolution of the reduced density matrix elements for the
single qubit to construct the reduced density matrix
$\hat{\rho}^T$ for the two-qubit system. In the standard product
basis $\mathcal{B}=\{\ket{1}\equiv\ket{11}, \ket{2}\equiv\ket{10},
\ket{3}\equiv\ket{01}, \ket{4}\equiv\ket{00} \}$, using
Eqs.~(\ref{singleevo}), (\ref{totalevo}) and (\ref{roA}), we
obtain the diagonal elements
\begin{eqnarray}\label{rototdiag}
\rho^T_{11}(t)&=&\rho^T_{11}(0)P_t^2,\nonumber\\
\rho^T_{22}(t)&=&\rho^T_{22}(0)P_t+\rho^T_{11}(0)P_t(1-P_t),\nonumber\\
\rho^T_{33}(t)&=&\rho^T_{33}(0)P_t+\rho^T_{11}P_t(1-P_t),\nonumber\\
\rho^T_{44}(t)&=&1-[\rho^T_{11}(t)+\rho^T_{22}(t)+\rho^T_{33}(t)],
\end{eqnarray}
and the non-diagonal elements
\begin{eqnarray}\label{rototnodiag}
\rho^T_{12}(t)&=&\rho^T_{12}(0)P_t^{3/2},\quad \rho^T_{13}(t)=\rho^T_{13}(0)P_t^{3/2},\nonumber\\
\rho^T_{14}(t)&=&\rho^T_{14}(0)P_t,\quad \rho^T_{23}(t)=\rho^T_{23}(0)P_t,\nonumber\\
\rho^T_{24}(t)&=&\sqrt{P_t}[\rho^T_{24}(0)+\rho^T_{13}(0)(1-P_t)],\nonumber\\
\rho^T_{34}(t)&=&\sqrt{P_t}[\rho^T_{34}(0)+\rho^T_{12}(0)(1-P_t)],
\end{eqnarray}
and $\rho^T_{ij}(t)=\rho^{T*}_{ji}(t)$, where $\rho^T(t)$ is a
hermitian matrix. In order to follow the entanglement dynamics of
the bipartite system, we use Wootters concurrence \cite{wootters}.
This is obtained from the density matrix $\hat{\rho}^T$ for qubits
A and B as $C_{\hat{\rho}^T}(t)=\mathrm{max}\{0,
\sqrt{\lambda_1}-\sqrt{\lambda_2}-\sqrt{\lambda_3}-\sqrt{\lambda_4}\}$,
where the quantities $\lambda_i$ are the eigenvalues of the matrix
$\zeta$
\begin{equation}
\zeta =\hat{\rho}^T (\sigma_y^A \otimes \sigma_y^B)\hat{\rho}^{T*}
(\sigma_y^A \otimes \sigma_y^B)\,,
\end{equation}
arranged in decreasing order. Here $\hat{\rho}^{T*}$ denotes the
complex conjugation of $\hat{\rho}^T$ in the standard basis and
$\sigma_y$ is the well-known Pauli matrix expressed in the same
basis. The concurrence varies from $C=0$ for a disentangled state
to $C=1$ for a maximally entangled state.

The form of Eqs.~(\ref{rototdiag}) and (\ref{rototnodiag}) is such
that one can study the entanglement evolution for any initial
state. We shall however restrict our analysis to the initial
entangled states
\begin{eqnarray}
\ket{\Phi}=\alpha\ket{01}+\beta\ket{10},\quad\ket{\Psi}=\alpha\ket{00}+\beta\ket{11},\label{istates}
\end{eqnarray}
where $\alpha$ is real, $\beta=|\beta|e^{i\delta}$ and
$\alpha^2+|\beta|^2=1$. For these two entangled states, the
initial total density matrix has an ``X'' structure \cite{yu5}
which is maintained during the evolution, as easily seen from
Eqs.~(\ref{rototdiag}) and (\ref{rototnodiag}). In particular the
concurrence, for the initial states of Eq.~(\ref{istates}), is
given by
\begin{eqnarray}
C_{\Phi}(t)&=&\mathrm{max}\{0,2|\rho^T_{23}(t)|-2\sqrt{\rho^T_{11}(t)\rho^T_{44}(t)}\},\nonumber\\
C_{\Psi}(t)&=&\mathrm{max}\{0,2|\rho^T_{14}(t)|-2\sqrt{\rho^T_{22}(t)\rho^T_{33}(t)}\},
\end{eqnarray}
and using Eqs.~(\ref{rototdiag}) and (\ref{rototnodiag}) we obtain
\begin{eqnarray}
C_{\Phi}(t)&=&\mathrm{max}\{0,2\sqrt{1-\alpha^2}
P_t\alpha\},\nonumber\\
C_{\Psi}(t)&=&\mathrm{max}\{0,2\sqrt{1-\alpha^2}P_t[\alpha-\sqrt{1-\alpha^2}(1-P_t)]\}.\nonumber\\\label{concu}
\end{eqnarray}
The time behavior of the concurrences $C_{\Phi}$ and $C_{\Psi}$ as
a function of $\alpha^2$ and the dimensionless quantity $\gamma_0
t$ are plotted for $P_t$ given by Eq.~(\ref{p1}) in Figs.~1 and 2
(for $\lambda/\gamma_0=0.1$). This is a condition that can be well
within the current experimental capabilities. In fact, cavity QED
experimental configurations have been realized using Rydberg atoms
with lifetimes $T_\textrm{at}\approx30\textrm{ms}$, inside
Fabry-Perot cavities with quality factors $Q\approx
4.2\times10^{10}$ corresponding to cavity lifetimes
$T_\textrm{cav} \approx 130\textrm{ms}$ \cite{kuhr}; these values
correspond to $2\lambda/\gamma_0 \approx 0.2$

\begin{figure}
\includegraphics[height=0.18\textheight, width=0.3\textwidth]{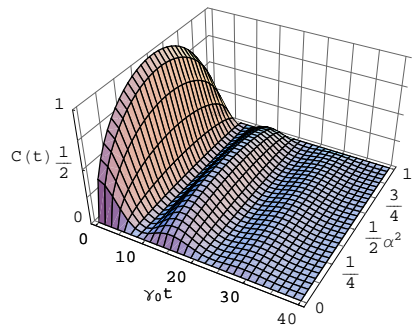}
\caption{\label{fig1}Concurrence for the initial state
$\alpha\ket{01}+\beta\ket{10}$ as a function of the dimensionless
quantity $\gamma_0t$ and $\alpha^2$, in a realistic CQED condition
($\lambda=0.1\gamma_0$).}
\end{figure}
\begin{figure}
\includegraphics[height=0.18\textheight, width=0.3\textwidth]{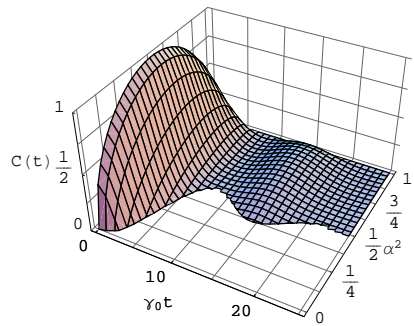}
\caption{\label{fig2}Concurrence for the initial state
$\alpha\ket{00}+\beta\ket{11}$ as a function of the dimensionless
quantity $\gamma_0t$ and $\alpha^2$, in a realistic CQED condition
($\lambda=0.1\gamma_0$).}
\end{figure}
Fig.~1 shows that, in the non-Markovian regime, the concurrence
$C_{\Phi}$ periodically vanishes according to the zeros of the
function $P_t$, with a damping of its revival amplitude. This
behavior is evidently different from Markovian, where in contrast
$C_\Phi$ decays exponentially and vanishes only asymptotically
\cite{santos}. The Markovian decay rate is however larger than the
initial non-Markovian one, as shown in Fig.~3 for the maximally
entangled case $\alpha^2=1/2$.
\begin{figure}
\includegraphics[height=0.15\textheight, width=0.27\textwidth]{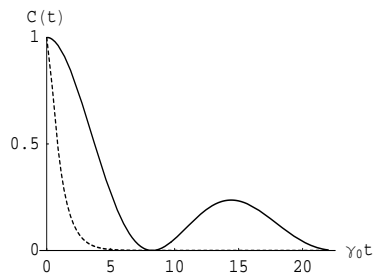}
\caption{\label{fig3}Concurrence for the initial state
$(\ket{01}+\ket{10})/\sqrt{2}$ as a function of the dimensionless
quantity $\gamma_0t$ in non-Markovian regime (solid line;
$\lambda=0.1\gamma_0$) and Markovian regime (dashed line;
$\lambda=5\gamma_0$).}
\end{figure}

Fig.~2 shows that the entanglement represented by $C_{\Psi}$ has a
similar behavior to $C_{\Phi}$ for $\alpha^2\geq 1/2$. In contrast
for $\alpha^2<1/2$ two ranges of parameter may be distinguished.
In one there is ESD because $C_{\Psi}$ vanishes permanently after
a finite time, similar to the Markovian case \cite{yu1,santos}. In
the second, revival of entanglement appears after periods of times
when disentanglement is complete. This behavior is more evident in
the plot of Fig.~4 obtained under stronger non-Markovian
conditions. This revival phenomenon is induced by the memory
effects of the reservoirs, which allows to the two-qubit
entanglement to reappear after a dark period of time, during which
the concurrence is zero.
\begin{figure}
\includegraphics[height=0.15\textheight, width=0.27\textwidth]{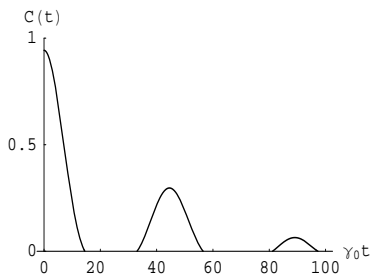}
\caption{\label{fig4}Concurrence for the initial state
$\alpha\ket{00}+\beta\ket{11}$ as a function of the dimensionless
quantity $\gamma_0t$ for $\alpha^2=1/3$, in strong coupling
($\lambda=0.01\gamma_0$).}
\end{figure}
This phenomenon of revival of entanglement after finite periods of
"entanglement death" appears to be linked to the
environment-single qubit non-Markovian dynamics. In this sense
this result differs from the revival of entanglement previously
obtained in the presence of interaction among qubits or because of
their interaction with a common reservoir \cite{tanas,sun,hamdou}.
The physical conditions examined here are moreover more similar to
those typically considered in quantum computation, where qubits
are taken to be independent and where qubits interact with
non-Markovian environments typical of solid state micro devices
\cite{vega}.

The above analysis can be easily extended to study entanglement
dynamics starting from different initial conditions and to take
into account finite temperature effects. In particular, starting
from a Werner state \cite{werner} one gets an entanglement
behavior structurally similar to that obtained in this letter for
the states of Eq.~(\ref{istates}) with ESD and entanglement
revival periods. The details of the evolution for this case and
finite temperature effects for non-Markovian dynamics will be
considered elsewhere.

In conclusion we have presented a procedure that allows in
principle to obtain the dynamics of a system of $N$ independent
bodies, each locally interacting with an environment, as long as
the single system dynamics is known. This procedure is valid for
any form of single body-environment interaction. It has been
applied to the case of two qubits interacting with the environment
where non-Markovian effects are present. In particular the model
described has been identified with a system made by two atoms each
in a high-$Q$ cavity. For this case the Hamiltonian dynamics of
the single qubit can be solved exactly and no problem about map
positivity arise. We have found that non-Markovian effects
influence the entanglement dynamics and may give rise to a revival
of entanglement even after complete disentanglement has been
present for finite time periods. This effect, arising for
completely independent systems, is only a consequence of the
non-Markovian behavior of the single qubit-reservoir dynamics.

These results show that entanglement dynamics may present facets
than one may not simply expect from single qubit dynamics and may
also lead to the possibility of recovering the entanglement
initially present.

\end{document}